\documentclass{eptcs}

\providecommand{\event}{DCM}
\providecommand{\volume}{??}
\providecommand{\anno}{2009}
\providecommand{\firstpage}{1}

 \usepackage{breakurl}
 \usepackage{epsfig}
\usepackage{amsfonts}
\usepackage{amssymb}
\usepackage{color}

\usepackage{amsmath}
\usepackage{prooftree}

\newcommand{\qqop}[1]{\mathrel{\makebox[2em]{$#1$}}}

\newcommand{\agr}{\quad\big|\quad}

\newcommand{\mycdot}{\!\cdot\!}
\newcommand{\CC}{\mathcal{C}}

\newcommand{\EE}{\mathcal{E}}

\newcommand{\RR}{\mathcal{R}}
\newcommand{\PP}{\mathcal{P}}
\newcommand{\Seq}{\mathcal{S}}
\newcommand{\XX}{\mathcal{X}}
\newcommand{\TV}{\mathcal{TV}}
\newcommand{\SV}{\mathcal{SV}}
\newcommand{\TT}{\mathcal{T}}

\newcommand{\SSq}{\mathcal{S}}

\newcommand{\VV}{\mathcal{V}}
\newcommand{\xx}{\widetilde{x}}
\newcommand{\yy}{\widetilde{y}}
\newcommand{\zz}{\widetilde{z}}

\newcommand{\ltrans}[1]{\xrightarrow{}}
\newcommand{\Ltrans}[1]{\Longrightarrow}
\newcommand{\Loop}[1]{\left(#1\right)^L}

\newcommand{\into}{\ensuremath{\,\rfloor}\,}
\newcommand{\pipe}{\ensuremath{\,|\,}}
\newcommand{\phole}{\square}

\newcommand{\ty}{{\tt t}}

\newcommand{\G}{\Gamma}
\newcommand{\pair}[2]{(#1,#2)}

\newcommand{\set}[1]{\{#1\}}
\newcommand{\ok}[2]{#1\bowtie #2}

\newcommand{\der}[4]{#1\vdash #2:\pair{#3}{#4}}
\newcommand{\build}[4]{~\vdash #1:#2; #3; #4}
\newcommand{\m}{{\sf m}}
\newcommand{\Th}{{\Theta}}

\newtheorem{theorem}{Theorem}[section]
\newtheorem{definition}[theorem]{Definition}

\newtheorem{lemma}[theorem]{Lemma}

\allowdisplaybreaks[1]

\newcommand{\Pe}{{\tt P}}
\newcommand{\R}{{\tt R}}
\newcommand{\E}{{\tt E}}
\newcommand{\T}{\Delta}
\newcommand{\three}[3]{(#1,#2)}
\newcommand{\derp}[5]{#1\vdash #2:\three{#3}{#4}{#5}}
\newcommand{\upper}[2]{#1\sqcup#2}
\newcommand{\derpST}[3]{#1\vdash #2:#3}
\newcommand{\excl}[1]{\E_{#1}}

\newcommand{\core}[1]{{\sf core}(#1)}

\title{A Type System for Required/Excluded Elements in CLS\thanks{This work was partly funded by the project BioBIT of the Regione Piemonte.}}
\author{Mariangiola Dezani-Ciancaglini
\institute{Dipartimento di Informatica, Universit\`a di Torino}
\email{dezani@di.unito.it}
\and
Paola Giannini
\institute{Dipartimento di Informatica, Universit\`a del Piemonte Orientale}
\email{giannini@mfn.unipmn.it}
\and
Angelo Troina
\institute{Dipartimento di Informatica, Universit\`a di Torino}
\email{troina@di.unito.it}
}

\def\titlerunning{A Type System for Required/Excluded Elements in CLS}
\def\authorrunning{M. Dezani-Ciancaglini, P. Giannini and A. Troina}

\begin{document}
\maketitle
\begin{abstract}
The calculus of looping sequences is a formalism for describing the
evolution of biological systems by means of term rewriting rules. We
enrich this calculus with a type discipline to guarantee
the soundness of reduction rules with respect to some
biological properties deriving from the requirement of certain elements, and the repellency of others.
As an example, we model a toy system where the repellency of a certain element is captured by our type system and forbids another element to exit a compartment.
\end{abstract}

\section{Introduction}

While the approach of biologists to describe biological systems by
mathematical means, allows them to reason on the behaviour of the
described systems and to perform quantitative simulations, such
modelling starts becoming more difficult both in specification and
in analysis when the complexity of the described systems increases.
This has become one of the main motivations for the application of
Computer Science formalisms to the description of biological
systems~\cite{RS02}. Other motivations can be found in the fact that
the use of formal means of Computer Science permits the application
of analysis methods that are practically unknown to biologists, such
as static analysis and model checking.

Many formalisms have either been applied to or have been inspired
from biological systems. The most notable are automata-based models
\cite{ABI01,MDNM00}, rewrite systems \cite{DL04,P02}, and process
calculi \cite{RS02,RS04,RPSCS04,Car05}. Models based on automata have the great advantage
of allowing the direct use of many verification tools such as model
checkers. On the other side, models based on rewrite systems usually allow describing biological
systems with a notation that can be easily understood by biologists.
However, automata-like models and rewrite systems present,
in general, problems from the point of view of compositionality.
Compositionality allows studying the behaviour of a system
componentwise, and is in general ensured by process calculi,
included those commonly used to describe biological systems.

In~\cite{BMMT06,BMMT06s,M07}, Milazzo et al. developed a new
formalism, called Calculus of Looping Sequences (CLS for short), for
describing biological systems and their evolution. CLS is based on
term rewriting with some features, such as a commutative parallel
composition operator, and some semantic means, such as bisimulations
\cite{BMMT06s,BMMT08}, which are common in process calculi. This
permits to combine the simplicity of notation of rewrite systems
with the advantage of a form of compositionality.

In chemistry, hydrophobicity is the physical property of a molecule
(known as a hydrophobe) that is repelled from a mass of water.
Hydrophobic molecules tend to be non--polar and thus prefer other
neutral molecules and non--polar solvents. Hydrophobic molecules in
water often cluster together forming micelles. From the other
perspective, water on hydrophobic surfaces will exhibit a high
contact angle (thus causing, for example, the familiar dew drops on
a hydrophobic leaf surface). Examples of hydrophobic molecules
include the alkanes, oils, fats, and greasy substances in general.
Hydrophobic materials are used for oil removal from water, the
management of oil spills, and chemical separation processes to
remove non-polar from polar compounds. Hydrophobicity is just an
example of repellency in Biochemistry. Other well known examples may
be found on the behaviour of anions and cations, or at a different
level of abstraction, in the behaviour of the rh antigen for the
different blood types.

As a counterpart, there may be elements, in nature, which always
require the presence of other elements (it is difficult to find a
lonely atom of oxygen, they always appear in the pair O$_2$).

In this paper we bring these aspects at their maximum limit, and, by
abstracting away all the phenomena which give rise/arise to/from
repellency (and its counterpart), we assume that for each kind of
element of our reality we are able to fix a set of elements which
are required by the element for its existence and a set of elements
whose presence is forbidden by the element.

Thus, we enrich CLS with a type discipline which allows to guarantee
the soundness of reduction rules with respect to some relevant
properties of biological systems deriving from the required and
excluded kinds of elements. The key technical tool we use is to
associate to each reduction rule the minimal set of conditions an
instantiation must satisfy in order to assure that applying this
rule to a ``correct'' system we get a ``correct'' system as well.

To the best of our knowledge~\cite{BMM07,ADT08} are the only papers which
study type disciplines for CLS. We generalise the proposal in~\cite{ADT08}
by focusing on the type disciplines for Present/Required/Excluded elements.

\section{The Calculus of Looping Sequences}\label{CLS_formalism}

In this section we recall the Calculus of Looping Sequences (CLS).
CLS is essentially based on term rewriting, hence a CLS model
consists of a term and a set of rewrite rules. The term is intended
to represent the structure of the modelled system, and the rewrite
rules to represent the events that may cause the system to evolve.

We start with defining the syntax of terms. We assume a possibly
infinite alphabet $\EE$ of symbols ranged over by $a,b,c,\ldots$.

\begin{definition}[Terms] \emph{Terms} $T$ and \emph{sequences} $S$ of {\em CLS} are
given by the following grammar:
\[
\begin{array}{lcl}
T\; & \qqop{::=} &S \agr \Loop{S} \into T \agr T \pipe T\\
S\; & \qqop{::=} &\epsilon \agr a  \agr S \cdot S
\end{array}
\]
 where $a$ is a generic element of $\EE$, and $\epsilon$ represents
the empty sequence. We denote with $\TT$ the infinite set of terms,
and with $\Seq$ the infinite set of sequences.
\end{definition}

In CLS we have a sequencing operator $\_\cdot\_$, a looping operator
$\Loop{\_}$, a parallel composition operator $\_\pipe\_$ and a
containment operator $\_\into\_$. Sequencing can be used to
concatenate elements of the alphabet $\EE$. The empty sequence
$\epsilon$ denotes the concatenation of zero symbols. A term can be
either a sequence or a looping sequence (that is the application of
the looping operator to a sequence) containing another term, or the
parallel composition of two terms.
By definition, looping and containment are always applied together,
hence we can consider them as a single binary operator $\Loop{\_}
\into \_$ which applies to one sequence and one term.

The biological interpretation of the operators is the following: the
main entities which occur in cells are DNA and RNA strands,
proteins, membranes, and other macro--molecules. DNA strands (and
similarly RNA strands) are sequences of nucleic acids, but they can
be seen also at a higher level of abstraction as sequences of genes.
Proteins are sequence of amino acids which usually have a very
complex three--dimensional structure. In a protein there are usually
(relatively) few subsequences, called domains, which actually are
able to interact with other entities by means of chemical reactions.
CLS sequences can model DNA/RNA strands and proteins by describing
each gene or each domain with a symbol of the alphabet. Membranes
are closed surfaces, often interspersed with proteins, which may
contain something. A closed surface can be modelled by a looping
sequence. The elements (or the subsequences) of the looping sequence
may represent the proteins on the membrane, and by the containment
operator it is possible to specify the content of the membrane.
Other macro--molecules can be modelled as single alphabet symbols, or
as short sequences. Finally, juxtaposition of entities can be
described by the parallel composition of their representations.

\begin{figure}[t]
\begin{center}
\begin{minipage}{0.98\textwidth}
\begin{center}
\psfig{figure=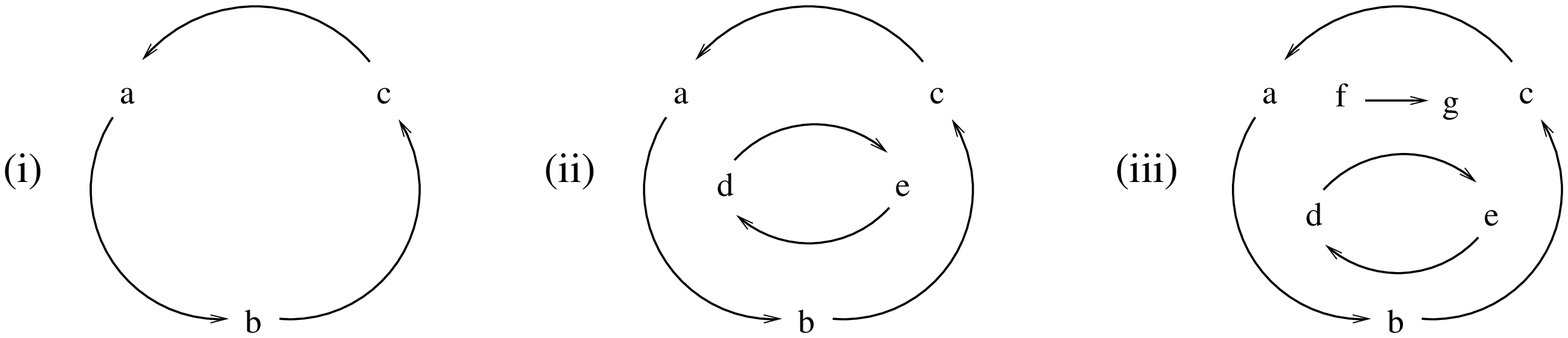, scale=0.43}
\end{center}
\caption{(i) represents $\Loop{a
\cdot b \cdot c}$; (ii) represents $\Loop{a
\cdot b \cdot c} \into \Loop{d \cdot e}$; (iii) represents $\Loop{a \cdot b
\cdot c} \into (\Loop{d \cdot e} \pipe f \cdot g)$.}
\label{fig:loop_seq_fig_CLS}
\end{minipage}
\end{center}
\end{figure}

Brackets can be used to indicate the order of application of the
operators, and we assume $\Loop{\_} \into \_$ to have precedence
over $\_\pipe\_$. In Figure~\ref{fig:loop_seq_fig_CLS} we show some
examples of CLS terms and their visual representation, using $\Loop{S}$ as a short-cut for $\Loop{S} \into \epsilon$.

In CLS we may have syntactically different terms representing the
same structure. We introduce a structural congruence relation to
identify such terms.

\begin{definition}[Structural Congruence] The structural
congruence relations $\equiv_S$ and $\equiv_T$ are the least
congruence relations on sequences and on terms, respectively,
satisfying the following rules:
\[
\begin{array}{c}
S_1 \cdot ( S_2 \cdot S_3 )
    \equiv_S ( S_1 \cdot S_2 ) \cdot S_3 \qquad
S \cdot \epsilon \equiv_S \epsilon \cdot S
    \equiv_S S
\\
S_1 \equiv_S S_2\ \mbox{ implies }\ S_1 \equiv_T S_2\ \mbox{ and }\
    \Loop{S_1} \into T \equiv_T \Loop{S_2} \into T\\
T_1 \pipe T_2 \equiv_T T_2 \pipe T_1 \qquad T_1 \pipe ( T_2 \pipe
T_3 ) \equiv_T (T_1 \pipe T_2) \pipe T_3 \qquad
T \pipe \epsilon \equiv_T T \\
\Loop{\epsilon} \into \epsilon \equiv_T \epsilon \qquad \Loop{S_1
\cdot S_2} \into T \equiv_T \Loop{S_2 \cdot S_1} \into T
\end{array}
\]
\end{definition}

Rules of the structural congruence state the associativity of
$\cdot$ and $\pipe$, the commutativity of the latter and the neutral
role of $\epsilon$. Moreover, axiom $\Loop{S_1 \cdot S_2} \into T
\equiv_T \Loop{S_2 \cdot S_1} \into T$ says that looping sequences
can rotate. In the following, for simplicity, we will use $\equiv$
in place of $\equiv_T$.

Rewrite rules will be defined essentially as pairs of terms, with
the first term describing the portion of the system in which the
event modelled by the rule may occur, and the second term describing
how that portion of the system changes when the event occurs. In the
terms of a rewrite rule we allow the use of variables. As a
consequence, a rule will be applicable to all terms which can be
obtained by properly instantiating its variables. Variables can be
of three kinds: two of these are associated with the two different
syntactic categories of terms and sequences, and one is associated
with single alphabet elements. We assume a set of term variables
$\TV$ ranged over by $X,Y,Z,\ldots$, a set of sequence variables $\SV$
ranged over by $\xx,\yy,\zz,\ldots$, and a set of element variables
$\XX$ ranged over by $x,y,z,\ldots$. All these sets are possibly
infinite and pairwise disjoint. We denote by $\VV$ the set of all
variables, $\VV = \TV \cup \SV \cup \XX$, and with $\rho$ a generic
variable of $\VV$. Hence, a pattern is a term that may include
variables.

\begin{definition}[Patterns] \emph{Patterns} $P$ and \emph{sequence patterns}
$SP$ of {\em CLS} are given by the following grammar:
\[
\begin{array}{lcl}
P\; & \qqop{::=} SP \agr \Loop{SP} \into P \agr P \pipe P \agr X\\
SP\; & \qqop{::=} \epsilon \agr a  \agr SP \cdot SP \agr \xx \agr x
\end{array}
\]
where $a$ is a generic element of $\EE$, and $X,\xx$ and $x$ are
generic elements of $\TV,\SV$ and $\XX$, respectively. We denote with
$\PP$ the infinite set of patterns.
\end{definition}

We assume the structural congruence relation to be trivially
extended to patterns. An \emph{instantiation} is a partial function
$\sigma : \VV \rightarrow \TT$. An instantiation must preserve the
kind of variables, thus for $X \in \TV, \xx \in \SV$ and $x \in \XX$
we have $\sigma(X) \in \TT, \sigma(\xx) \in \Seq$ and $\sigma(x) \in
\EE$, respectively. Given $P \in \PP$, with $P \sigma$ we denote the
term obtained by replacing each occurrence of each variable $\rho
\in \VV$ appearing in $P$ with the corresponding term
$\sigma(\rho)$. With $\Sigma$ we denote the set of all the possible
instantiations and, given $P \in \PP$, with $Var(P)$ we denote the
set of variables appearing in $P$. Now we define rewrite rules.

\begin{definition}[Rewrite Rules] A rewrite rule is a pair of patterns
$(P_1,P_2)$, denoted with $P_1 \! \mapsto \! P_2$, where $P_1,P_2
\in \PP$, $P_1 \not\equiv \epsilon$ and such that $Var(P_2)
\subseteq Var(P_1)$.
\end{definition}

A rewrite rule $P_1 \! \mapsto \! P_2$ states that a term $P_1
\sigma$, obtained by instantiating variables in $P_1$ by some
instantiation function $\sigma$, can be transformed into the term
$P_2\sigma$. We define the semantics of CLS as a transition system,
in which states correspond to terms, and transitions correspond to
rule applications.

We define the semantics of CLS by resorting to the notion of
contexts.

\begin{definition}[Contexts] \emph{Contexts} $C$ are defined as:
\[
C ::= \phole \agr C \pipe T \agr T \pipe C \agr \Loop{S} \into C
\]
where $T \in \TT$ and $S \in \SSq$. The context $\phole$ is called
the \emph{empty context}. We denote with $\CC$ the infinite set of
contexts.
\end{definition}

By definition, every context contains a single hole $\phole$. Let us
assume $C,C'\in \CC$. With $C[T]$ we denote the term obtained by
replacing $\phole$ with $T$ in $C$; with $C[C']$ we denote context
composition, whose result is the context obtained by replacing
$\phole$ with $C'$ in $C$. The structural equivalence is extended to
contexts in the natural way (i.e. by considering $\phole$ as a new
and unique symbol of the alphabet $\EE$).

Rewrite rules can be applied to terms only if they occur in a legal
context. Note that the general form of rewrite rules does not permit
to have sequences as contexts. A rewrite rule introducing a parallel
composition on the right hand side (as $a \mapsto b \pipe c$)
applied to an element of a sequence (e.g., $m \mycdot a \mycdot m$)
would result into a syntactically incorrect term (in this case $m
\cdot (b\pipe c) \cdot m$). To modify a sequence, a pattern
representing the whole sequence must appear in the rule. For
example, rule $a \mycdot \xx \mapsto a \pipe \xx$ can be applied to
any sequence starting with element $a$, and, hence, the term
$a\mycdot b$ can be rewritten as $a \pipe b$, and the term $a
\mycdot b \mycdot c$ can be rewritten as $a \pipe b \mycdot c$.

The semantics of CLS is defined as follows.

\begin{definition}[Semantics]
Given a finite set of rewrite rules $\RR$, the \emph{semantics} of
{\em CLS} is the least relation closed with respect to $\equiv$ and
satisfying the following rule:
\[
\prooftree
\begin{array}{c} P_1 \mapsto P_2 \in \RR \quad P_1\sigma \not\equiv
\epsilon \quad \sigma \in \Sigma \quad C\in \CC \end{array}
\justifies C[P_1\sigma] \ltrans{} C[P_2\sigma]
\endprooftree
\]
\end{definition}

As usual we denote with $\ltrans{}^*$ the reflexive and transitive
closure of $\ltrans{}$.

Given a set of rewrite rules $\RR$, the behaviour of a term $T$ is
the tree of terms to which  $T$ may reduce. Thus, a \emph{model} in
CLS is given by a term describing the initial state of the system
and by a set of rewrite rules describing all the events that may
occur.

\section{A Type Discipline for Required and Excluded
Elements}\label{tdpre}

We classify elements in $\EE$ with {\em basic types}.  Intuitively,
given a molecule represented by an element in $\EE$, we associate to
it a type \ty\ which specifies the kind of the molecule. We assume a
fixed typing $\G$ for the elements in $\EE$.

For each basic type \ty\ we assume to have a pair of sets of basic types
$(\R_{\ty},{\E_{\ty}})$, where $\ty\not\in\R_{\ty}\cup\E_{\ty}$ and
$\R_{\ty}\cap\E_{\ty}=\emptyset$, saying that the presence of
elements of basic type $\ty$ requires the presence of elements whose basic type
is in $\R_{\ty}$ and forbids the presence of elements whose basic type is
in $\E_{\ty}$. We consider only {\em local} properties: elements
influence each other if they are either in the same compartment or
they contain each other. 

The type system derives the set of types of patterns (and
therefore also terms), checking that the
constraints imposed by the required and excluded sets are
satisfied. Types are pairs $\three\Pe\R\E$: where \Pe\ is the set of
basic types of {\em present} elements (at the top level of a pattern), \R\
is the set of basic types of {\em required} elements (that should still be
added to the pattern to represent a \emph{correct}
system). The set of {\em excluded} elements is implicitly given by
$\excl{\Pe}=\bigcup_{\ty\in\Pe}\E_{\ty}$. 

Types are well formed, and pair of types are compatible, if their
constraints on required and excluded elements are not contradictory;
compatible types can be combined.
\begin{definition}[Auxiliary definitions]
\begin{itemize}
\item A type $\three\Pe\R\E$ is {\em well formed} if
$\Pe\cap\excl{\Pe}=\Pe\cap\R=\R\cap\excl{\Pe}=\emptyset$.
 \item
Well formed types {\em $\three\Pe\R\E$ and $\three{\Pe'}{\R'}{\E'}$
are compatible} (written
$\ok{\three\Pe\R\E}{\three{\Pe'}{\R'}{\E'}}$) if
\begin{itemize}
  \item $\excl{\Pe}\cap\Pe'=\excl{\Pe}\cap\R'=\emptyset$, and
  \item $\excl{\Pe'}\cap\Pe=\excl{\Pe'}\cap\R=\emptyset$.
\end{itemize}
\item
Given two compatible types $\three\Pe\R\E$ and
$\three{\Pe'}{\R'}{\E'}$ we define their \emph{conjunction}
$\upper{\three{\Pe}{\R}{\E}}{\three{\Pe'}{\R'}{\E'}}$ by \[\upper{\three{\Pe}{\R}{\E}}{\three{\Pe'}{\R'}{\E'}}=\three{\Pe\cup\Pe'}{(\R\cup\R')\setminus(\Pe\cup\Pe')}{\E\cup\E'}.\]
\end{itemize}
\end{definition}
Basis are defined by:
\[
\T\qqop{::=}\emptyset\agr\T,x:\three{\set\ty}{\R_\ty}\E\agr\T,\eta:\three\Pe\R\E
 \]
where $\eta$ denotes a sequence or term variable. A basis $\T$ is
{\em well formed} if all types in the basis are well formed.

We check the safety of terms, sequences and more generally patterns
using the typing rules of Figure \ref{rpre}. It is easy to verify
that, if we start from well-formed basis, then in a derivation we
produce only well-formed basis and well-formed types. Note that
terms and sequences are typable from the empty context.
\begin{figure}[h]
\[
\begin{array}{c}
\begin{array}{ccccc}
\derp{ \T,\rho:\three\Pe\R\E}{\rho}\Pe\R\E&\qquad& \derp
\T\epsilon\emptyset\emptyset\emptyset&\qquad& \prooftree
a:\ty\in\G \justifies \derp{ \T}a{\set\ty}{\R_{\ty}}{\E_{\ty}}
\endprooftree
\end{array}
\\ \\
\prooftree \derp{ \T}{SP}\Pe\R\E\quad \derp{
\T}{SP'}{\Pe'}{\R'}{\E'}\quad
 \ok{\three{\Pe}{\R}{\E}}{\three{\Pe'}{\R'}{\E'}}\justifies \derpST{
\T}{SP\mycdot SP'}{\upper{\three\Pe\R\E}{\three{\Pe'}{\R'}{\E'}}}
\endprooftree \\ \\
\prooftree \derp{ \T}{P}\Pe\R\E\quad \derp{
\T}{P'}{\Pe'}{\R'}{\E'}\quad
\ok{\three{\Pe}{\R}{\E}}{\three{\Pe'}{\R'}{\E'}}
  \justifies \derpST{ \T}{P\pipe
P'}{\upper{\three\Pe\R\E}{\three{\Pe'}{\R'}{\E'}}}
\endprooftree \\ \\
\prooftree \derp{ \T}{SP}\Pe\R\E\quad \derp{
\T}{P}{\Pe'}{\R'}{\E'}\quad
\ok{\three{\Pe}{\R}{\E}}{\three{\Pe'}{\R'}{\E'}}\mbox{ and
}\R'\subseteq\Pe
\justifies \derp{ \T}{\Loop{SP} \into
P}\Pe{\R\setminus\Pe'}\E
\endprooftree
\end{array}
\]
\caption{Typing Rules}\label{rpre}
\end{figure}
All the rules are trivial except for the last one which types
looping sequences. In this rule we can put a pattern $P$ inside a
looping sequence $SP$ only if all the types required from $P$ are
provided by $SP$. This is because if $P$ gets inside a compartment
(represented by the looping sequence) it cannot interact any more
with the
environment.

Given a context we define the possible types of terms that may fill
the hole in the context.
\begin{definition}[Typed Holes]\label{d:typedContext} Given a
context $C$, and a well-formed type $\three{\Pe}{\R}{}$, the type $\three{\Pe}{\R}{}$ is {\em OK} for the context $C$
if
$\derpST{X:\three{\Pe}{\R}{}}{C[X]}{\three{\Pe'}{\emptyset}{}}$ for
some $\Pe'$.
\end{definition}
The above notion guarantees that filling a context with a term we
obtain a correct system (whose type is well formed and whose
requirements are completely satisfied). It is to this kind of terms
that we are interested in applying reduction rules.

Note that there may be more than one type $\three{\Pe}{\R}{}$ such
that $\three{\Pe}{\R}{}$ is  OK for the context $C$.

We can classify reduction rules according to the types we can derive for the right hand sides of the rules.
\begin{definition}[$\T$-$\three{\Pe}{\R}{}$-Reduction Rules]\label{typedrule}A rule
$P_1 \mapsto P_2$ is a {\em $\T$-$\three{\Pe}{\R}{}$-reduction rule}
if $\derpST{\T}{P_2}{\three{\Pe}{\R}{}}$.
\end{definition}
An instantiation $\sigma$ {\em agrees} with a basis $\T$
(notation $\sigma\in\Sigma_\T$) if $\rho:\three\Pe\R\E\in \T$
implies $\derpST{}{\sigma(\rho)}{\three\Pe\R\E}$.

We can safely
apply a rule to a typed term only if the instances of the pattern on the
right hand side of the rule has a type that is {\em OK} for the
context. More formally:
\begin{definition}[Typed Semantics]\label{defTypedSem}
Given a finite set of rewrite rules $\RR$, the \emph{typed
semantics} of {\em CLS} is the least relation closed with respect to
$\equiv$ and satisfying the following rule:
 \[
 \prooftree
\begin{array}{c} P_1 \mapsto P_2 \in \RR \text{ is a
$\T$-$\three{\Pe}{\R}{}$-reduction rule}\qquad P_1\sigma \not\equiv
\epsilon \\ \sigma \in \Sigma_\T \qquad C\in \CC\qquad
\three\Pe\R\E\ \ \mbox{is OK for}\ \  C
\end{array} \justifies
 C[P_1\sigma] \Ltrans{\T} C[P_2\sigma]
    \endprooftree
\]
\end{definition}
As expected, reduction preserves typing, in the sense that the
obtained term is still typable, but the new type can have a
different set of present elements, while the set of required
elements is always empty. This choice makes possible typing creation
and degradation of elements.
\begin{theorem}
If $\derp{}{T}\Pe\emptyset\E$ and $T \Ltrans{\T} T'$, then
$\derp{}{T'}{\Pe'}\emptyset\E$ for some $\Pe'$.
\end{theorem}
We can infer the OK relation between types and contexts and which
rules are  $\T$-$\three{\Pe}{\R}{}$-reduction rules by using the
machinery of principal typing \cite{well02}. In this way we can
decide the applicability of the reduction rules for the typed
semantics. This is the content of the remaining part of this
section.

We convene that for each variable $x\in\XX$ there is an  {\em e-type
variable} $\varphi_x$ ranging over basic types, and for each
variable $\eta\in \TV\cup \SV$ there are two variables $\phi_\eta$,
$\psi_\eta$ (called {\em p-type variable} and {\em r-type variable})
ranging over sets of basic types. Moreover we convene that $\Phi$
ranges over formal unions and differences of sets of basic types, e-type variables and p-type
variables, and $\Psi$ ranges
over formal unions and differences of sets of basic types and r-type variables.\\
A \emph{basis scheme} $\Th$ is a mapping from atomic variables to
their e-type variables, and from sequence and term variables to
pairs of their p-type variables and r-type variables:
\[\Th\qqop{::=}\emptyset\agr\Th,x:\varphi_x\agr\Th,\eta:\pair{\phi_\eta}{\psi_\eta}.\]
The rules for inferring principal typing use judgements of the shape:

\[\build {P}{\Th}{\pair \Phi \Psi}{\Xi}\]
\noindent where $\Th$ is the {\em principal basis} in which $P$ is
well formed, ${\pair \Phi \Psi}$ is the {\em principal type} of $P$,
and $\Xi$ is the set of constraints that should be satisfied.
Figure~\ref{ptir} gives these inference rules.
\begin{figure}
\begin{center}

$\build\epsilon \emptyset{\pair\emptyset\emptyset}\emptyset$ \qquad
\prooftree a:\ty\in\G
\justifies
\build a {\emptyset} {\pair\ty{\R_\ty}}\emptyset
\endprooftree
 \qquad
$\build x {\set{x:\pair{\varphi_x}\Psi}} {\pair{\varphi_x}\Psi}\set{\Psi=\R_{\varphi_x}}$

\medskip

$\build \eta
{\set{\eta:{\pair{\phi_\eta}{\psi_\eta}}}}{\pair{\phi_\eta}{\psi_\eta}}
\emptyset$

\medskip

\centerline{\prooftree \build {SP}{\Th}{\pair \Phi \Psi}{\Xi}\;\qquad
\build {SP'}{\Th'}{\pair {\Phi'}{\Psi'}}{\Xi'} \justifies \build {SP
\mycdot SP'}{\Th \cup \Th'}{\upper{\pair\Phi \Psi}{\pair {\Phi'}{\Psi'}}}
{\Xi \cup\Xi'\cup\set{\ok{\pair\Phi \Psi}{\pair {\Phi'}{\Psi'}}}}
\endprooftree}

\bigskip

\centerline{\prooftree \build {P}{\Th}{\pair \Phi \Psi}{\Xi}\;\qquad
\build {P'}{\Th'}{\pair {\Phi'}{\Psi'}}{\Xi'} \justifies \build {P \;|\;
P'}{\Th \cup \Th'}{\upper{\pair\Phi \Psi}{\pair {\Phi'}{\Psi'}}}
{\Xi \cup\Xi'\cup\set{\ok{\pair\Phi \Psi}{\pair {\Phi'}{\Psi'}}}}
\endprooftree}

\bigskip

\centerline{\prooftree \build {SP}{\Th}{\pair \Phi \Psi}{\Xi}\;\qquad
\build {P}{\Th'}{\pair {\Phi'}{\Psi'}}{\Xi'} \justifies \build
{\Loop{SP} \into P}{\Th \cup \Th'}{\pair
{\Phi}{\Psi\setminus\Phi'}}{\Xi \cup \Xi'\cup\set{\ok{\pair\Phi \Psi}{\pair {\Phi'}{\Psi'}},\Psi'\subseteq\Phi}}
\endprooftree}

\end{center}
\caption{Inference Rules for Principal Typing}\label{ptir}
\end{figure}

Soundness and completeness of our inference rules can be stated as
usual. A {\em type mapping} maps e-type variables to basic types,
p-type variables and r-type variables to sets of basic types. A type
mapping $\m$ {\em satisfies} a set of constraints $\Xi$ if  all
constraints in $\m(\Xi)$ are satisfied.
\begin{theorem}[Soundness of Type Inference]
If $\build {P}{\Th}{\pair \Phi \Psi}{\Xi}$ and $\m$ is a type
mapping which satisfies $\Xi$, then
$\der{\m(\Th)}{P}{\m(\Phi)}{\m(\Psi)}$.
\end{theorem}
\begin{theorem}[Completeness of Type Inference]
If $\der{\T}{P}\Pe\R$, then $\build {P}{\Th}{\pair \Phi \Psi}{\Xi}$
for some  $\Th$, $\pair \Phi \Psi$, $\Xi$ and there is a type
mapping $\m$ that satisfies $\Xi$ and  such that
$\T\supseteq\m(\Th)$, $\Pe=\m(\Phi)$, $\R=\m(\Psi)$.
\end{theorem}
We put now our inference rules at work in order to decide
applicability of typed reduction rules. We first characterize by
means of principal typing the OK relation  and the classification of
reduction rules.

Notably for deciding the OK relation it is not necessary to consider
the whole context, but only the part of the context which influences
the typing of the hole. The key observation is that the typing of a
term inside two nested looping sequences does not depend on the
typing of the terms outside the outermost looping sequence. We call
{\em core of the context} this part. More formally:
\begin{definition}
The {\em core of the context} $C$ (notation $\core C$) is defined by:
 \begin{itemize}
\item $\core C=C\quad$ if $\quad C\equiv \phole\pipe T_1\quad $ or $\quad C\equiv
\Loop{S_1} \into (\phole\pipe T_1)\pipe T_2$;
  \item $\core {C}=C_2\quad$ if $\quad C=C_1[C_2]\quad\mbox{where}\quad C_2\equiv \Loop{S_2} \into (\Loop{S_1} \into (\phole\pipe T_1)\pipe
  T_2)$.
\end{itemize}
\end{definition}
Remark that $\sf core$ is always unambiguously defined, since every
context can be split in a unique way into one of the
three shapes of the previous definition.
\begin{lemma}[OK Relation] Let the context $C$ be such that $\derpST{}{C[T]}{\three{\Pe_0}{\emptyset}{}}$ for
some $T,\Pe_0$.
A type $\pair\Pe\R$ is OK for $C$ if and only if the type
mapping $\m$ defined by
\begin{enumerate}
\item$\m(\phi_X)=\Pe$,
\item $\m(\psi_X)=\R$,
\end{enumerate}
satisfies the set of constraints \[\Xi\cup\set{\Psi=\emptyset\mid \text { if }\phi_X \text{ or }\psi_X\text{ occurs in }\Psi},\]
where $\build {\core{C}[X]}{\set{X:\pair{\phi_X}{\psi_X}}}{\pair \Phi
\Psi}{\Xi}$.
\end{lemma}
It is easy to check that if $\core{C}\equiv \Loop{S_2} \into
(\Loop{S_1} \into (\phole\pipe T_1)\pipe T_2)$, and
$\der{}{T_1}{\Pe_1}{\R_1}$, $\der{}{S_1}{\Pe'_1}{\R'_1}$,
$\der{}{T_2}{\Pe_2}{\R_2}$,  $\der{}{S_2}{\Pe'_2}{\R'_2}$, then we
have to verify the following six constraints:
\begin{itemize}
\item $\ok{\pair{\phi_X}{\psi_X}}{\pair{\Pe_1}{\R_1}}$
\item $\ok{\pair{\Pe_1'}{\R_1'}}{(\upper{\pair{\phi_X}{\psi_X}}{\pair{\Pe_1}{\R_1}})}$
\item $((\psi_X\cup\R_1)\setminus(\phi_X\cup\Pe_1))\subseteq \Pe'_1$
\item $ \ok{\pair{\Pe'_1}{\R'_1\setminus(\phi_X\cup\Pe_1)}}{\pair{\Pe_2}{\R_2}}$
\item $\ok{\pair{\Pe'_2}{\R'_2}}{(\upper{\pair{\Pe'_1}{\R'_1\setminus(\phi_X\cup\Pe_1)}}{\pair{\Pe_2}{\R_2}})}$
\item $(((\R'_1\setminus(\phi_X\cup\Pe_1))\cup\R_2)\setminus(\Pe_1'\cup\Pe_2))\subseteq \Pe'_2.$
\end{itemize}
The set of constraints is smaller when the core context
has one of the simpler shapes.
\begin{lemma}[Classification of Reduction Rules] A rule
$P_1 \mapsto P_2$ is a $\T$-$\three{\Pe}{\R}{}$-reduction rule if
and only if the type mapping $\m$ defined by \begin{enumerate}\item
$\m(\varphi_x)=\ty$ if $\T(x)=\pair{\set{\ty}}{\R_\ty}$, \item
$\m(\phi_\eta)=\Pe'$ if $\T(\eta)=\pair{\Pe'}{\R'}$,  \item
$\m(\psi_\eta)=\R'$ if $\T(\eta)=\pair{\Pe'}{\R'}$, \end{enumerate}
satisfies the set of constraints $\Xi\cup\set{\Phi=\Pe,\Psi=\R}$,
where $\build {P_2}{\Th}{\pair \Phi \Psi}{\Xi}$.
\end{lemma}
The previous two lemmas imply the following theorem which gives the
desired result.
\begin{theorem}[Applicability of Reduction Rules]
Let \[\build {P_2}{\Th}{\pair \Phi \Psi}{\Xi}\quad\quad\text{ and
}\quad\build {\core{C}[X]}{\set{X:\pair{\phi_X}{\psi_X}}}{\pair
{\Phi'} {\Psi'}}{\Xi'}.\] Then the rule $P_1 \mapsto P_2$ can be
applied to the term $C[P_1\sigma]$ such that
$\derpST{}{C[P_1\sigma]}{\three{\Pe}{\emptyset}{}}$ for some $\Pe$
if and only if the type mapping $\m$ defined by
\begin{enumerate}\item $\m(\varphi_x)=\ty$ if $\sigma(x):\ty\in\G$,
\item $\m(\phi_\eta)=\Pe'$ if $\der{}{\sigma(\eta)}{\Pe'}{\R'}$,
\item $\m(\psi_\eta)=\R'$ if $\der{}{\sigma(\eta)}{\Pe'}{\R'}$,
\end{enumerate} satisfies the set
of constraints
$\Xi\cup\Xi'\cup\set{\Phi=\phi_X,\Psi=\psi_X}\cup\set{\Psi'=\emptyset\mid \text { if }\phi_X \text{ or }\psi_X\text{ occurs in }\Psi'}$.
\end{theorem}
Note that - after fixing the reduction rules - the sets of constraints
for typing the r.h.s. of these rules can be evaluated once for all.
Instead, the sets of constraints for typing the core contexts need
to be evaluated at every application of a reduction rule. Luckily
these sets of constraints includes at most six constraints. The mapping  $\m$
is immediate from the derivation of a type for $P_1\sigma$. Finally,
the checking that $\m$ satisfies a set of constraints
requires only some substitutions.

\section{Examples}\label{examples}
We start showing the properties of our type system by modelling an
example of two molecules repelling each other. As we have seen, one
might model repellency in our framework via the set $\E_\ty$.

Namely, if molecule $a$, of basic type $\ty$, is a repellent for
molecule $b$, of basic type $\ty'$ (and viceversa), we will have
that $\E_\ty=\{\ty'\}$ and $\E_{\ty'}=\{\ty\}$. Note that this does
not mean that $a$ and $b$ cannot be present in the same term,
actually they should just be contained in two different
compartments. In fact, the term
$$
T=a \pipe \Loop{m} \into b
$$
with $m$ of basic type $\ty''$, where $\E_{\ty''}=\emptyset$ and
$\R_\ty=\R_{\ty'}=\R_{\ty''}=\emptyset$, is typed by the pair
$(\{\ty,\ty''\},\emptyset)$ by the following derivation, where
$\G=\set{a:\ty,b:\ty',m:\ty''}$:
\[
\prooftree {
 \prooftree a:\ty\in\G \justifies \derp{
}a{\set\ty}{\emptyset}{\E_{\ty}}
\endprooftree
\quad
 \prooftree{
\prooftree m:\ty''\in\G \justifies \derp{
}m{\set{\ty''}}{\emptyset}{\E_{\ty''}}
\endprooftree
\quad \prooftree b:\ty'\in\G \justifies \derp{
}b{\set{\ty'}}{\emptyset}{\E_{\ty'}}
\endprooftree \quad\ok{(\{\ty''\},\emptyset)}{(\{\ty'\},\emptyset)}}
 \justifies \derpST{
}{\Loop{m} \into b}{(\{\ty''\},\emptyset)}
\endprooftree \quad\ok{(\{\ty''\},\emptyset)}{(\{\ty\},\emptyset)}
 }
\justifies \derpST{ }{a \pipe \Loop{m} \into
b}{(\{\ty,\ty''\},\emptyset)}
\endprooftree
\]
As we can see the term is well typed, since $a$ and $b$ are in two
different compartments. The element $m$ could also be replaced by any sequence which does not repel elements of types $\ty,\ty'$ and whose
(well formed) type is $\three{\Pe}{\R}{}$ where
$\ok{\three{\Pe}{\R}{}}{(\{\ty'\},\emptyset)}$ and
$\ok{\three{\Pe}{\R}{}}{(\{\ty\},\emptyset)}$, that is
$\Pe\cap\{\ty,\ty'\}=\emptyset$ and $\R\cap\{\ty,\ty'\}=\emptyset$.
Moreover, if the term is at top level, then $\R$ should be $\emptyset$.

Consider now the rule
$$
R_1: \Loop{\xx} \into (X \pipe b) \mapsto b \pipe \Loop{\xx} \into X
$$
which moves the element $b$ outside the compartment. Such a rule
could not be applied in $T$ cause it will result in the term $a\pipe
b \pipe\Loop{m}\into \epsilon$, which is not well typed since
$(\{\ty,\ty',\ty''\},\emptyset)$ is not well formed.
Indeed, following Definition~\ref{defTypedSem},
$P_1=\Loop{\xx} \into (X \pipe b)$, $P_2 = b \pipe \Loop{\xx} \into
X$, $\sigma(\xx)=m$, $\sigma(X)=\epsilon$ and $C= a\pipe \phole$.
Moreover, while $C[P_1\sigma]$ is well typed, the same does not hold
for $C[P_2\sigma]$.

This does not mean the rule $R_1$ can never be applied. In fact, if
we consider the term
$$
T'=a \pipe \Loop{m} \into b \pipe \Loop{m}\into \epsilon
$$
and we add the rule
$$
R_2: a \pipe \Loop{\xx}\into X \mapsto \Loop{\xx}\into (a \pipe X)
$$
we still have that rule $R_1$ cannot be applied to $T'$ (same
reasons as before). Moreover, if we consider the context $C=\Loop{m}
\into \epsilon \pipe \phole$, $P_1=a \pipe \Loop{\xx}\into X$, and $P_2=\Loop{\xx}\into (a \pipe X)$,
that is we try to move $a$ in the compartment containing $b$, we
cannot apply the rule $R_2$, since $\sigma(\xx)=m$ and
$\sigma(X)=b$, and therefore $P_2\sigma$ is not well typed. However,
we can apply rule $R_2$ to $T'$ if we consider $C= \Loop{m} \into b
\pipe \phole$, $P_1=a \pipe \Loop{\xx}\into X$, and
$P_2=\Loop{\xx}\into (a \pipe X)$, that is moving $a$ in the
compartment that does not contain $b$, since $\sigma(\xx)=m$ and
$\sigma(X)=\epsilon$, and we have that $T'\equiv C[P_1\sigma]$ has
type $(\{\ty,\ty''\},\emptyset)$ and $T''=C[P_2\sigma]=\Loop{m}\into
a \pipe \Loop{m} \into b$ has type $(\{\ty''\},\emptyset)$ - both
$C[P_1\sigma]$ and $C[P_2\sigma]$ are well typed. Now, we can apply
rule $R_1$ to $T''$, bringing $b$ outside the compartment. The
context of the rule application will be $C=\Loop{m}\into a \pipe
\phole$ and the patterns $P_1=\Loop{\xx} \into (X \pipe b)$, $P_2 =
b \pipe \Loop{\xx} \into X$ where $\sigma(\xx)=m$ and
$\sigma(X)=\epsilon$. Both $C[P_1\sigma]$ and $C[P_2\sigma]$ are
well typed and the resulting term will be
$$
T'''=b\pipe\Loop{m}\into a \pipe \Loop{m}\into \epsilon.
$$

We now show an application for the set of required elements in our typing system.
The idea is to model the absorption of a given compound $c$ by a cell. The absorption
is promoted by a receptor $r$ which should be present in the surface of the cell.
We can model the effect of the absorption by using a different symbol (thus a different type)
for the compound when it enters the cell, namely $c'$. The basic types of $c$, $r$ and $c'$ are,
respectively, $\ty$, $\ty'$ and $\ty''$. We also assume that $\E_\ty=\E_{\ty'}=\E_{\ty''}=\R_\ty=\R_{\ty'}=\emptyset$. The requirement for $\ty''$ (modelling the type of the compound inside a cell)
should be, instead, the presence of the receptor on the membrane surface. We can model this
condition with the set $\R_{\ty''}=\{\ty'\}$. Thus, with these basic types, we can model the rule for the absorption of the protein as:
$$
R: c \pipe \Loop{\xx}\into X  \mapsto \Loop{\xx}\into (X \pipe c')
$$
without imposing explicitly the presence of the receptors on the patterns of the rule.

Actually, our type system guarantees that such a rule cannot be
applied to the term $c \pipe \Loop{m}\into \epsilon$, while it is
applicable to the term $c \pipe \Loop{m\cdot r}\into \epsilon$.

In a sense, the role of the promoter (the receptor) is modelled intrinsically on the type of the compound
brought inside the cell; its properties become transparent for the rewrite rules
and controlled by the type system.

\section{Conclusions}
This paper is a first step toward the application of typing to the
safety of system transformations which model biological phenomena.
We focused on disciplines deriving by the requirement/exclusion of
certain elements, and used the type system to describe how
repellency could be modelled. We would like to underline that in
nature it is not easy to find elements which completely exclude or
require other elements. Our abstraction, however allows us to deal
with a simple qualitative model, and to observe some basic
properties of biological systems. A more detailed analysis, could
also deal with quantities. In this case, typing is useful in
modelling quantitative aspects of CLS semantics on the line of
\cite{BMMTT08}. In particular, in \cite{DGT09}, we show how types
can be used to model repellency also by quantitative means, that is
slowing down undesired interactions.

As a future work, we plan to investigate type disciplines
assuring different properties for CLS and to apply this approach to
other calculi for describing evolution of biological systems, in
particular to P systems.

An interesting application of this model may also abstract from
biological phenomena. In a sense, the composition of a context $C$
with a term $T$ which satisfies $C$ may represent the agreement on a
\emph{contract} between $C$ and $T$. Namely, if $C$ satisfies
$(\Pe,\R)$ and $T$ has type $(\Pe,\R)$, then $T$ offers to $C$, via
the elements in $\Pe$, everything that is \emph{required} by $C$,
viceversa, $C$ has to satisfy $T$'s request $\R$; modelling, in a
sense, the fact that $C$ and $T$ mutually agree with each other.

Finally, modelling biological phenomena in CLS, especially for
biologists, could be made more intuitive with the use of a graphical
interface, that would check most of the syntactical details, so that
the modeller could focus on the conceptual aspects of the
formalization.

\paragraph{Acknowledgements.} We thank the referees for their helpful comments.
The final version of the paper improved due to their suggestions.

\end{document}